\documentclass[runningheads]{llncs}
\usepackage{hyperref}
\usepackage{graphicx}   % for including graphics
\usepackage{tabularx}
\usepackage{booktabs}
\usepackage{subfigure}

\begin{document}
\title{Transforming Software Development with Generative AI: Empirical Insights on Collaboration and Workflow}% \thanks{Supported by organization x.}}
\titlerunning{Transforming Software Development with GenAI}
\author{Rasmus Ulfsnes\inst{1}\orcidID{0000-1111-2222-3333} \and Nils Brede Moe\inst{1}\orcidID{0000-0003-2669-0778} \and
Viktoria Stray\inst{1,2}\orcidID{0000-0002-6032-2074} \and
Marianne Skarpen\inst{3}\orcidID{0009-0005-5575-316X}
 } 
\authorrunning{R. Ulfsnes et al.}
\institute{SINTEF, Strindvegen 4, 7030 Trondheim, Norway\\
\email{\{rasmus.ulfsnes, nils.b.moe\}@sintef.no}\\
 \and
Department of Informatics, University of Oslo, Oslo, Norway\\
\email{stray@uio.no}\\
\and
NTNU, Trondheim, Norway \\
\email{marskarp@stud.ntnu.no    }
}

%\url{http://www.sintef.no}
%
\maketitle              % typeset the header of the contribution
\begin{abstract}
Generative AI (GenAI) has fundamentally changed how knowledge workers, such as software developers, solve tasks and collaborate to build software products. Introducing innovative tools like ChatGPT and Copilot has created new opportunities to assist and augment software developers across various problems. We conducted an empirical study involving interviews with 13 data scientists, managers, developers, designers, and frontend developers to investigate the usage of GenAI. Our study reveals that ChatGPT signifies a paradigm shift in the workflow of software developers. The technology empowers developers by enabling them to work more efficiently, speed up the learning process, and increase motivation by reducing tedious and repetitive tasks.  Moreover, our results indicate a change in teamwork collaboration due to software engineers using GenAI for help instead of asking co-workers which impacts the learning loop in agile teams. 

\end{abstract}

\section{Introduction}

There is a growing trend among companies to adopt digitalization and engage in digital transformation\cite{ulfsnes_technology_2023}. This transformation process requires technologies as software, data   and artificial intelligence \cite{bosch_digital_2021}, forcing a shift in the use of strategic frameworks \cite{stray_how_2022}, and new ways of developing technology for highly skilled employees with intelligent technology \cite{ulfsnes_technology_2023}. 

Technology for assisting developers with writing code, particularly using Integrated Development Environments (IDEs) is not a new concept \cite{murphy_how_2006,kersten_using_2006}. The task of autocompletion of code, and generating of test, and various other tasks has been particularly interesting for software, as natural language  matches quite well as the software code is hypothesized to be a natural language \cite{hindle_naturalness_2016}. Subsuquently this hypothesis has led to lots of research on Artificial Intelligence (AI) for software engineering \cite{svyatkovskiy_pythia_2019,talamadupula_applied_2021}. With the introduction of Generative Artificial Intelligence (GenAI) -  a type of artificial intelligence (AI), both the software development processes and tooling have started to change fast. Further, GenAI can revolutionize software development by automating repetitive tasks, improving code quality, enhancing collaboration, providing data-driven insights, and ultimately accelerating the development lifecycle \cite{sun_investigating_2022,muller_drinking_2022}. There is a growing research into how to use Copilot \cite{Ross_Martinez_Houde_Muller_Weisz_2023}, or Generative AI systems such as Chat GPT \cite{White_Hays_Fu_Spencer-Smith_Schmidt_2023}, and its capability to automate software engineering tasks \cite{Melegati_Guerra_2023}. However, good tooling is not enough. 

In order for a company to succeed with software product development, well-working teams and good processes are key. Software engineering is a social activity that is focused on close cooperation and collaboration between all team members \cite{mens_social_2019} and across teams in the organization \cite{berntzen2023responding}. Therefore, it is important to note that while AI has great potential, it also comes with challenges \cite{bender_dangers_2021} such as ethical considerations, data privacy concerns, the need for skilled professionals to handle the technology within software teams, and a potential change in the team dynamics. However, research on team dynamics is lacking. In order to understand the effects on team-dynamics however we also need to consider the individual work practices, to grasp the effects on a team level. This paper explores how are software engineer work practices transformed, and potential impact on the transformation on collaboration.

 We have interviewed 13 data scientists, managers, developers, designers, and frontend developers to investigate how they use GenAI technology and how their workday has changed. Finally, we discuss how this technology might affect software development teamwork.

\section{Related work}

\subsection{Productivity and work satisfaction}

Competing for talents requires a conscious effort to offer an attractive workplace \cite{moe_attractive_2023}. Further, the ability to balance the need and nature of the workdays for different team members is directly related to the outcome of the product \cite{tkalich_toward_2022}. For software developers, Meyer et al. \cite{meyer_today_2019} outline a framework that describes what makes a good workday. There are three main factors: value creation, efficient use of time, and perception. Value creation is about whether or not the developers feel they are creating something, and the factor has six sub-factors. 
The second factor, efficient use of time, has two sub-factors, meeting expectations and the ability to work focused. In essence, the assessment of a workday being good or bad is largely influenced by the expectations for the day. For example, if one anticipates a day filled with meetings, the day can be considered good even if most of the time is spent in meetings. However, if one hopes for a day of focused work and the day is filled with meetings, the day is perceived as a bad workday. Coworker interruptions were specifically described as negatively influencing developers' ability to focus or work as planned, although being able to help a co-worker was generally considered positive and rewarding. Lastly, perception is about how they perceive their own productivity. 

Developer satisfaction and work productivity are related, therefore they need to be key considerations for software companies \cite{forsgren_space_2021}. More productive developers may be more satisfied, and more satisfied developers may be more productive. Autonomy, being able to complete tasks, and technical skills all affect productivity. By introducing new technology like GenAI, a team member's productivity may be positively affected. At the same time work culture and team collaboration are important for job satisfaction. An increased reliance on tools like GenAI may enhance individual productivity while inadvertently reducing inter-team interactions, ultimately affecting long-term job satisfaction and collective productivity. Introducing GenAI in software teams is therefore a balancing act.

\subsection{Software development, knowledge work and technology}
Software engineering requires the input and consolidation of various information to produce code \cite{murphy_how_2006}. Furthermore, with the advent of DevOps with its increase in speed of delivering continuously \cite{fitzgerald_continuous_2015}, increased accessibility to third-party libraries and frameworks, the process of software engineering has shifted from being about understanding the computer and the programming language towards understanding how to compose relevant libraries and frameworks, with applicable testing. 

Software development encompasses more than just programming and teamwork, it also involves actively seeking knowledge online and in knowledge management systems \cite{dingsoyr_what_2009}, conducting testing and code reviewing \cite{florea2019global}, and taking advantage of software such as Integrated Development Environments. IDEs has been a researched topic for quite some time \cite{murphy_how_2006} as well as how online resources \cite{chatterjee_finding_2020,li_debugging_2022} aid and enhance the development process, both in speed and quality across varying ranges of experience. Through the use of IDEs, software engineers have gotten access to capabilities for refactoring, debugging, source repositories, third-party plugins\cite{murphy_how_2006} and auto-completion of code \cite{bruch_learning_2009}. In addition, developers need to browse through a plethora of different files in existing software solutions in order to get a grasp of how changes to the code need to be implemented \cite{kersten_using_2006}. %Furthermore, there is a clear difference between how novice and experienced 
%Even though new 

More recently, better auto-completion methodologies and technologies have been introduced to provide more context-relevant suggestions using statistical methods \cite{bruch_learning_2009}, and the 
naturalness of software \cite{hindle_naturalness_2016} is a great target for utilizing natural language processing and generative AI. Github Copilot has shown promising effects for assisting developers in writing code, assisting with test writing \cite{muller_drinking_2022,bird_taking_2022}. Early studies on knowledge work show that generative AI are able to disseminate knowledge previously shown as tacit\cite{brynjolfsson_generative_2023} and dramatically increase both quality and production\cite{dellacqua_navigating_2023}. Both studies show that the effects are most noticeable for the lowers skilled workers while higher skilled workers have a lower increase in prodcution and quality.

\subsection{Teams, knowledge sharing and performance}
Software product development is done in teams 
\cite{liu_modeling_2020}, therefore, the success of software development depends significantly on team performance. Today the premise is that software teams should be autonomous or self-managed \cite{ravn2022team}. In their review, Dickinson and McIntyre \cite{dickinson_conceptual_1997} identified and defined seven core components of teamwork. Using these components and their relationships as a basis, they proposed the teamwork model that is used in this work. The model consists of a learning loop of the following basic teamwork components: communication, team orientation, team leadership, monitoring, feedback, backup, and coordination. Later Moe et al. \cite{moe_teamwork_2010} used this model to explain agile teamwork. The introduction of GenAI is likely to affect these teamwork components.

\section{Research Method and Analysis}
This study was conducted in the context of two research programs on software development processes, where several companies introduced Generative Artificial Intelligence (GenAI) in their product development process. GenAI, especially those built on LLMs, is a new phenomenon that has not been previously studied. Due to the uncertain nature of the phenomenon, we chose an exploratory multi-case study \cite{yin_case_2018}. We selected our informants using snowball sampling \cite{biernacki_snowball_1981} in Slack asking for subjects that  used GenAI for a wide range of activities. As of the tools used, our studies found that ChatGPT and GitHub Copilot were the most common for code and text, while some reported that they used Midjourney and DALL-E 2 for image creation.

\subsection{Data Collection and Analysis}
 We interviewed 13 people, as shown in Table \ref{tab:informants}. The informant group had a broad range of roles: data scientists, managers, developers, designers, and front-end developers. Based on a literature review, we developed a semi-structured interview guide. Questions included: \textit{How do you use GenAI services?} and \textit{Which effects do you get from using them?}.  The interviews were done by the first, third, fourth, and fifth authors to spread out any subjective biases.  

 The analysis was divided into two cycles of coding as suggested by Saldana \cite{saldana_coding_2013}, with the third and first author conducting a combination of descriptive and initial coding \cite{corbin_basics_2014} on the first eight interviews. Then the third and first authors had a discussion about the emerging categories and themes. This led to a revised interview guide. Finally, the last five interviews were conducted. 
 
 After the last round of interviews, the fourth author performed a descriptive coding of the remaining interviews. At the same time, the first author performed a second cycle \cite{saldana_coding_2013} focused coding of all interviews. Then, the observed themes and categories were merged through mutual workshops and discussions with all authors.

\begin{table}[t]
    \small
    \caption{Data sources}
    \label{tab:informants}
    \begin{tabularx}{\textwidth}{c X c } 
        \toprule
        ID & Role & Work experience in years\\
        \midrule
        I1  & Developer \& Team lead & 5 \\
        I2  & Developer & 5 \\
        I3  & Technical Strategy Consultant \& Director & 30 \\
        I4  & Tech Lead \& Machine learning engineer & 6  \\
        I5  & Principal Engineer and Enterprise Architect & 15  \\
        I6  & Data Science Manager & 6 \\
        I7  & Developer & 10 \\
        I8  & CTO & 25 \\
        I9  & Director & 30 \\
        I10  & Developer & N/A \\
        I11  & Designer & 13 \\
        I12  & Designer & 3 \\
        I13  & Developer & 3 \\
        \bottomrule
    \end{tabularx}
\end{table}
\section{What is Generative AI Used for in Software Development?}
Software development consists of a number of activities, ranging across multiple roles in cross-functional teams. When using source repository systems or IDEs, the use case is often clear. However, the use of GenAI takes on a much more individual form. There are currently no standards or norms for how, when, and for what purpose you should apply GenAI to, and employees in software-intensive organizations are using it based on their own preferences. GenAI tools for a wide range of activities. See Table \ref{tab:gai_activites} for an overview of such activities.

The type of GenAI activities and utilization depends on individual preferences, the task to be solved, and the user's role in the organization. Developers typically use GenAI when working with the source code, while managers use it for, e.g. organizing workshops or creating content for PowerPoint presentations.

\subsection{Asking for assistance when stuck}
%\textbf{Asking for assistance when stuck}.  
When a person was stuck on a particular problem or did not know how to proceed, ChatGPT was used as an assistant or fellow team member, where interacting with it using chat could help a person solve complex problems or get increased progress. For non-technical problems, this could be a case of writer's block, formulations, or when they are zoning out: \emph{"For me, the main thing is to get unstuck, whether I am struggling with writer's block or formulations, just by interacting with ChatGPT and getting an immediate response is something else."} 

This highlights that there is an effect of just having the chat window open, and getting feedback without interrupting others. Formulating the problem to ChatGPT made it easier to keep focus on the task, helped on the thought processes, and helped see the problem in a new light. Developers referred to such interaction with ChatGPT as rubber-ducking. The idea of rubber-ducking is to explain the problem one seeks to solve to an inanimate object (e.g. a rubber duck), in an attempt to achieve a deeper understanding of the problem and a potential solution through the process of explaining it to someone (or something) using natural language.

Using GenAI as a sparring partner was both faster than asking human colleagues and also took away the feeling of disturbing them in their work. Further, being able to formulate the question as you would to a \emph{human} felt easier than the alternate Google search, where you need to consider the specific keywords, and what results they can give you.

\begin{table}[tb] \centering \caption{List of GenAI Activities} \label{tab:gai_activites} \begin{tabularx}{\textwidth}{p{4.2cm} >{\raggedright\arraybackslash}X}
        \toprule
        Activity & Description\\
        \midrule
        Asking for assistance when stuck & When stuck on a particular task, GenAI can help getting out of the slump. \\
        Learning & GenAI provide an interactive way of learning new things. \\
        Creating a virtual environment for a product & By asking GenAI to provide an simulated/virtual environment to learn and test products. \\
        PowerPoint and email writing & GenAI is useful for helping with writing text for email, powerpoint. \\
        Non-technical boilerplate & Providing a boilerplate for how to get started with powerpoint, workshops. \\
        Boilerplate code & GenAI can provide boilerplate code that acts as a skeleton for further development. \\ 
        Working with existing code & GenAI are used for refactoring, adding small features, making code more robust, converting code, debugging, writing tests, Search Engine Optimization (SEO). \\
        \bottomrule
    \end{tabularx}
\end{table}

\subsection{Learning}
GenAI provided more opportunities and avenues for engagement in learning compared to reading books, using Google, or watching videos. One approach was to engage interactively with the chat, assigning a role to the AI like "act as a tester", and then engaging with that persona to learn about testing. They could then ask that persona to explain a particular topic like they were doing testing for the first time.
%\begin{quote}
    
\emph{"And then I continue. And I notice that I learn much faster this way. Because it's like having a personal tutor. Where you can ask yourself questions. And then I always have to double-check."}
%\end{quote}

The same approach was also applied when working on code, where using new features, or when working on areas that the informants were not that knowledgeable about, a data scientist explained how to use the technology to learn more about programming: %\begin{quote}
\emph{"Almost as if I were asking someone much more skilled, like a developer in this case."}
%\end{quote}

\subsection{Virtual environments}
Creating a virtual environment for a product was used to develop software for a trading platform. They engaged with the Chat GPT and asked it to simulate that it was a stock exchange. They then provided the Chat GPT with information about which stocks could be traded at the exchange and asked it to simulate different trading scenarios. This provided a novel way for the informant to understand the intricacies of a stock exchange. Further, this means that the Chat GPT had the context for the particular trading platform the informants were interested in.
   
\emph{"And then I said, "now I'm going to make an application out of this in such-and-such language." And so it has the context for everything while I kept asking it further questions."} 
This integrated approach to both understanding a domain, also then produced the relevant context that ChatGPT could use to generate relevant code. 

\subsection{Copywriting}
Maybe not surprisingly, Chat GPT was used to assist in copywriting text, especially useful when integrated into the tool the informant was using, getting live feedback; this was particularly useful for persons that were not native or fluent in English or Norwegian. However, some experienced that GenAI was not very useful for email and text writing for two main reasons, there was a significant overhead in engaging sufficiently with ChatGPT to create emails, and the quality did not get better. 
%\begin{quote}

\emph{"I think I have asked it to write emails, but for me it is just faster to write it myself. The formulation was better though."}
%\end{quote}      

\subsection{Boilerplating code and text}
Getting started with a relatively novel coding project in any company requires quite a lot of boilerplate code; this type of code does not add functionality relevant to the business case but is required to get the project up and running with the necessary declarations and structures. Both back-end and front-end developers used ChatGPT to create tailored boilerplate code: 

%\begin{quote}
    \emph{"If I have a task, to create a list of tricks with something, and thumbs up and thumbs down on each element, for instance. I often start by describing what I want to ChatGPT. Then, it writes the code for me."}
%\end{quote}
GenAI was also used for repetitive or tedious non-technical tasks. For example, managers and architects stated to use ChatGPT to consolidate text used for production of bid to customers. One manager explained,  
%\begin{quote}
\emph{"A bid I would normally have spent a lot of time in writing, I only spent 20 minutes on. Previously I would have spent a lot of time, looking for previous bids, adapting it and merging it. 
It is terrible to say it out loud [laughing], as this kind of is in someway reducing the need for my work. This type of work is what the company pays me to do."}
%\end{quote}

Another example is when you are building up technical specifications and technical architectures where the style of the text is quite consistent but the content varies between use cases. Or getting feedback on emails, and getting a headstart on the writing of the text.   

\subsection{Working with existing code}

This was the most common activity among software developers. GenAI was used on many different tasks, ranging from refactoring or simplifying code, code review, translating code from one programming language to another, and simply explaining the code. Testing was also well-suited for GenAI utilization, given its repetitive nature, GenAI was used to create numerous tests for the code. The informants also noted that the generated tests sometimes accounted for scenarios and test cases that they themselves had not thought of.

%\begin{quote}
    \emph{"It was mostly a matter of thinking up all the things that could go wrong and creating unit tests for them. And that's where CoPilot was brilliant, as it came up with things that could go wrong that I had never thought of."}
%\end{quote}

%Nytt kapittel 

\section{How and why do we interact with GenAI?}
In the previous section, we described how GenAI is used for a variety of activities. In the studied companies GenAI was becoming an integrated part of their daily work, and most explained that they used GenAI daily, or "all the time". 

Among the study participants we found two styles of interaction: \emph{simple dialogue} and \emph{advanced dialogue extended with prompt engineering}. 
The interaction style depended on the work context and types of problems to be solved. Table \ref{tab:mode_ben_ch} contains an overview of the effects and drawbacks from interacting with GenAI. 

\begin{table}[t]
    \centering
    \caption{Use of GenAI - Benefits and Challenges}
    \label{tab:mode_ben_ch}
    \begin{tabularx}{\textwidth}{X X X}
       \toprule
       Mode  & Benefits & Challenges \\ 
       \midrule

       Simple %& In Q\& A mode, the users will ask questions without providing context and then use that answer with minimal dialogue. 
       & 
            \begin{itemize}
              \item Asking questions is easier than searching
              \item ChatGPT immediately provides an (almost) usable result
               \item More efficient
               \item More fun
               \item More time to learn
               \item Increased motivation and work satisfaction
            \end{itemize}
        & \begin{itemize}           
            \item Input cleaning 
            \item Lack of tool integration
           \item Lack of information after 2021
           \item Output needs to be worked on
           \item Culturally biased output
        \end{itemize} \\
      Advanced (Simple and Prompt engineering) &
      %In when prompting the users will provide context, and explain how they want GenAI to respond to the users' input. & 
      \begin{itemize}
           \item Interactive learning
           \item More precise answers
           \item Ability to take on different roles
           \item Pair-prompt engineering
           \item ChatGPT as advisor - rubberducking
       \end{itemize} & \begin{itemize}
           \item Limited context ability
           \item Less pair-programming
           \item Still requires other people when the complexity increases
            \item Prompt engineering requires competence
       \end{itemize} \\
       \bottomrule
    \end{tabularx}
\end{table}

\subsection{Effects}
By spending less time on manual and repetitive tasks, the improved productivity brought more enjoyment, motivation and fun to the work. The repetitive tasks were seen as menial, and not particularly mentally challenging. Further, as time was freed up, more time could be spent on creative and challenging tasks. Moreover, engaging with the GenAI itself  was experienced as fun and increased the motivation to experiment with different applications of the new technology. 

Interacting through dialogue with ChatGPT increased engagement. It was experienced as a more "natural" engagement then searching for answers to problems on Google. Having a dialogue with ChatGPT was also described as faster than concocting the necessary string of Google search keywords. Moreover, ChatGPT responds immediately with the, assumed, correct answer to the question, while googling often required additional steps, vetting the correct site on the search page, entering the particular page, and analysing the web page for the potential answer to the question. One informant was so conscious about speed that they deliberately chose GPT v3.5 over v4 in certain cases (at the time of our data collection v3.5 was faster than v4) , where the precision and quality of the v3.5 answer was assumed to be sufficient. 

GenAI's utility also extended beyond quick and precise answers, with informants reporting a freedom in interacting with an artificial tool rather than having to deal with the social considerations involved in asking a team member.

%\begin{quote}
    \emph{"Yeah, so you don't need to be too polite either. You don't have to have the correct phrasing or anything. You can just throw something out, I feel. Then you can get an answer, and if it's not quite right, you can refine the question again."}
%\end{quote}

The threshold of asking ChatGPT was significantly lower than asking another person or in a Slack channel. This threshold for asking colleagues could potentially be high in a busy work environment, as one does not wish to interrupt an already-busy colleague. Further, you get feedback immediately, while it might take a while to get feedback on Slack.

\subsection{Challenges}

While there are many positive benefits of using GenAI, there are also challenges. The elusiveness of data confidentiality, data policy, and sensitivity meant that everyone was acutely conscious about which data to input into the chat interface. This made the work process somewhat awkward, requiring cleansing and anonymization of the text being sent into ChatGPT. Developers in companies using open-source technology were less lenient in protecting code than those in companies with internal code repositories. Moreover, the general lack of tool integration meant that there was a substantial amount of copy-paste to move text and code between different windows. One developer using Copilot X reported that the integration in the IDE meant that the code could be autocompleted, and explained by ChatGPT in a seamless process, which reduced their work immensely:
%\begin{quote}
    \emph{"I think it would have been easier to adopt a GenAI tool if I had used something like Copilot. Because then it would have been, in a way, integrated into the workflow."}
%\end{quote}

With regards to the output, all informants noted that the content produced by GenAI, regardless of tool, seldom represented a final product and typically required further refinement to be applicable in a real-world context. The general attitude from the interviewees was that they expected the output to be wrong. 

Regarding technological development, which is characterized by a rapid pace and a increasing number of available libraries and technology, the cut off date for ChatGPT's training data in September 2021 represented a significant drawback, were the error-rate was annoyingly high.

One architect creating project startup documents, experienced that ChatGPT was culturally biased towards how more hierarchical companies would perform activities in a project. This meant that ChatGPT had to be prompted with specific information regarding the methodology and project practices: 

%\begin{quote}
    \emph{"You kind of have to trick it into the right context if it's (GenAI) going to be part of agile processes."}
%\end{quote}

\subsection{Prompt engineering}
Several interviewees talked about how the quality of their prompt affected the quality of the response, and how the use of prompt engineering techniques like contextualizing the problem, using personas, etc. to guide and steer the dialogue with ChatGPT. Prompt engineering was applied to all the activities in table \ref{tab:gai_activites}. One informant explained asking ChatGPT to create a description of the most critical code reviewer in the world. They then told ChatGPT to act like this description while reviewing the codein a pull request. Another more technical aspect was telling ChatGPT to act like an SQL database to test queries. The effect of using prompt engineering was seen as a matter of precision and quality, thus reducing the time spent on working on modifying the output. One explained the usage of prompt engineering as follows:

%\begin{quote}
    \emph{"It's like putting up fences on the bowling lane and then narrowing it down even more. It can almost only go one way, and that's a strike."}
%\end{quote}

An important prompt engineering technique was assigning different roles to ChatGPT for the same question to get more than one perspective or answer on a problem. One explained,

%\begin{quote}
    \emph{"I want you to respond like a wealth manager," "I want you to respond like a friend," or "like a so-and-so..." And then you get different answers."}
%\end{quote}

Using prompt engineering while writing code was described as feeling similar to programming with a partner. The flip side was that the developers mentioned that they were doing less pair programming as they were getting the wanted rubber ducking effect from using GenAI. 

\section{Discussion}
One of the notable consequences of integrating GenAI tools into software development tasks is a visible shift in collaborative communication dynamics. Some of the informants appear to have a growing inclination to consult AI-driven solutions for issues and tasks they previously discussed with their human colleagues. This shift can have dramatic effects on the team's ability to perform. According to Liu et al., \cite{liu_modeling_2020}, a team's ability to perform is highly dependent on the knowledge sharing of the team. This implies that reducing knowledge sharing by replacing this with Generative should be observed. This opposes the findings by Brynjolfsson et al. \cite{brynjolfsson_generative_2023}, where knowledge dissemination between high-skilled and low-skilled workers in customer service. These findings point to a significant difference between work done by teams in software development and individual work in customer service. However, we find that individuals become more efficient and save time, which they spend on more rewarding tasks. 

\begin{figure}[ht]
    \centering
    \subfigure[Regular learning loop]{
        \includegraphics[width=0.8\linewidth]{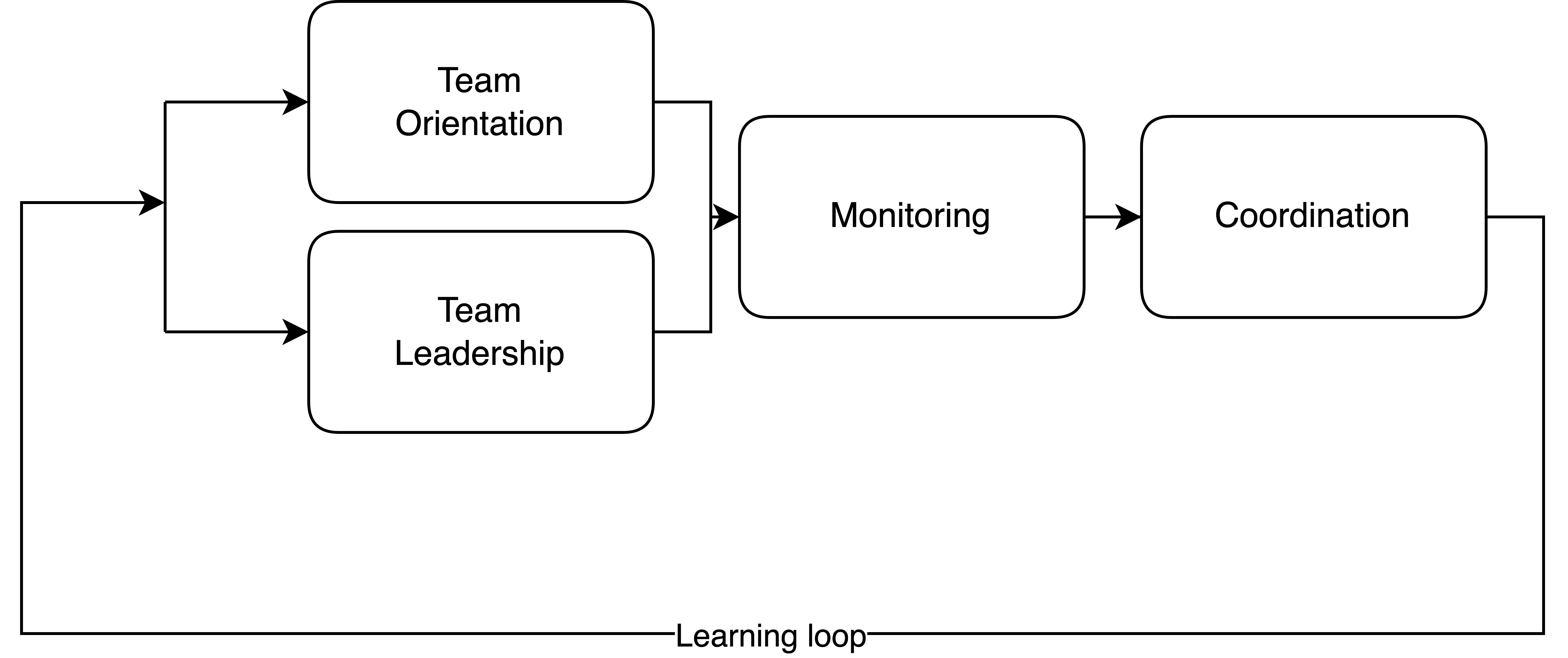}
        \label{fig:subfig_a}
    }
    %\quad % Adds some space between the figures
    \subfigure[Disrupted learning loop]{
        \includegraphics[width=0.8\linewidth]{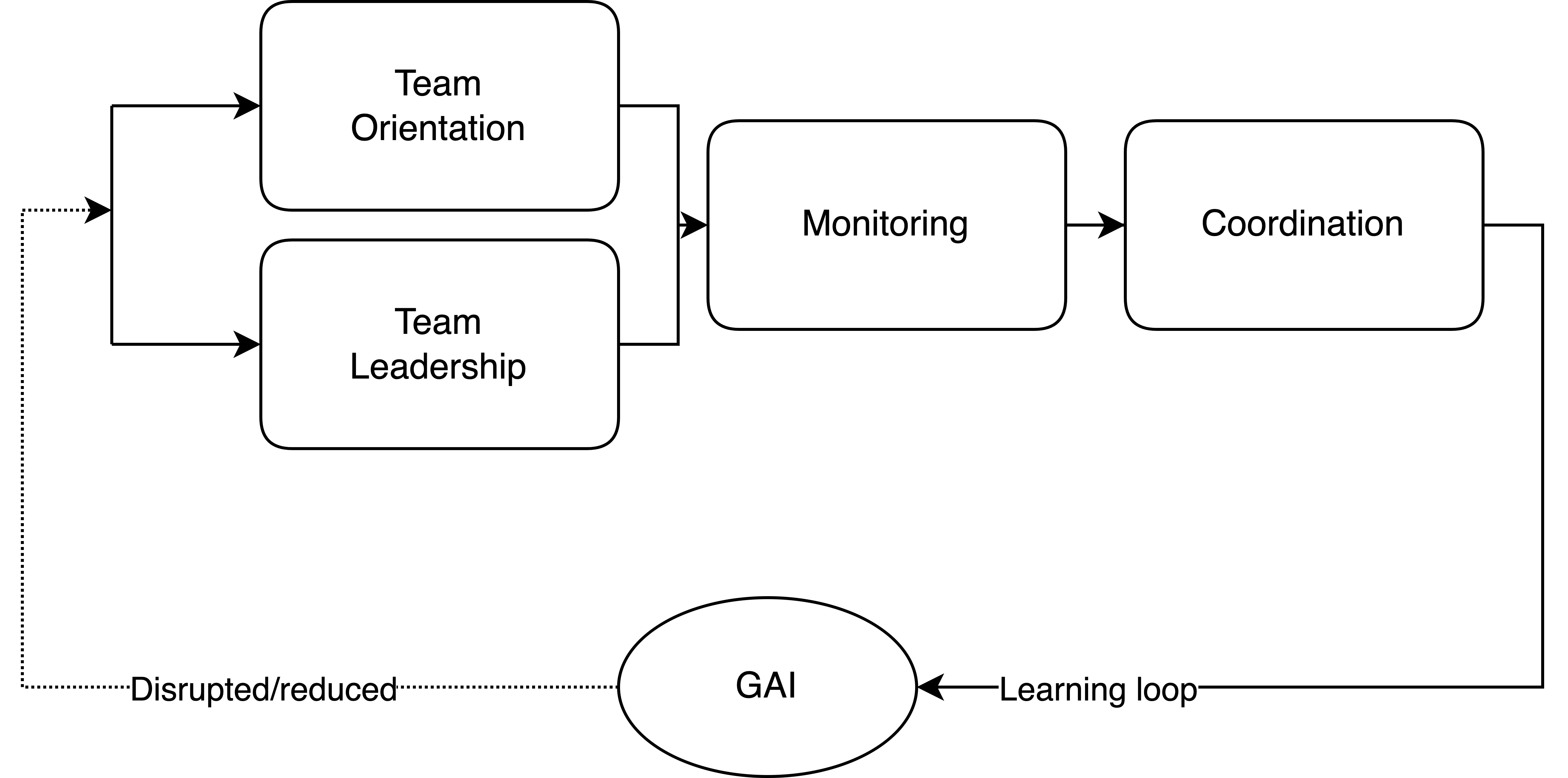}
        \label{fig:subfig_b}
    }
    \caption{Regular and disrupted learning loops.}
    \label{fig:combined}
\end{figure}

If team members reduce their interactions in favor of focusing on individual tasks, a phenomenon known as an isomorphic team structure may emerge. The advantages of this structure are that it is organizationally simple, allows many tasks to be completed in parallel, and can clearly define and understand task responsibilities. However, the effect of such a structure is that the developers focused on their own modules and often created their own plans and made their own decisions. In addition, problems are seen as personal, individual goals are more important than team goals, and team members become less aware of what others are doing and get less support and help from others \cite{moe_teamwork_2010}. In a good working team, learning is a continuous feedback, see Figure \ref{fig:subfig_a}. By introducing GenAI; this loop will be disrupted, or reduced, Figure \ref{fig:subfig_b}, thus reducing teamwork performance\cite{moe_teamwork_2010}. Additionally, this can also contribute to making persons less satisfied as helping others is a key factor for good workdays \cite{meyer_today_2019}. 

This model posits that the incorporation of GenAI in software development may disrupt the established learning loop.  Such disruption will subsequently affect individual and team performance in software development. While it is anticipated that GenAI might enhance individual performance by streamlining tasks, there is a concurrent risk of diminishing overall team performance. 

However, everything is not dark; as multiple informants noted, they had an increasing amount of knowledge sharing on how to use GenAI in their work and context. Notably, the practice of "pair prompt engineering" has emerged, akin to the concept of pair programming. This approach facilitates knowledge sharing \cite{tkalich_pair_2023} both on-site and when working remotely. This can thus involve a shift in how the programmers program, creating yet another abstraction layer for code production. 

\section{Concluding remarks}

In essence, GenAI serves a dual purpose: making everyday tasks more efficient and reigniting creative thinking for leaders and developers. By automating the production of routine code snippets and related tasks, these tools enable programmers to focus on higher-level conceptualization and innovation, resulting in enhanced productivity and code quality. This is similar to findings by Meyer et al. \cite{meyer_today_2019}, where good workdays are understood as days where they feel productive and are able to work focused. In addition, similar to Brynjolfsson et al. where there is knowledge dissemination through the GenAI \cite{brynjolfsson_generative_2023}, we observe that there are data scientists using GenAI for coding purposes and frontend developers getting assistance in back-end development. Both programmers and leaders acknowledged the potential of generative AI in freeing up valuable time and cognitive resources that could be better allocated to more creative and complex problem-solving tasks. 

This is reported to enhance both efficiency and enjoyment. By automating the production of routine code snippets and documentation, these GenAI enabled programmers to focus on higher-level conceptualization and other more complicated tasks, resulting in higher reported productivity. Most informants reported that AI reduced the amount of time developers spent on projects. 

%\subsection{The New Era of Rubberducking?}
%Potential for future studies
\section*{Acknowledgments}
This work was supported by the Research Council of Norway grants, 321477, and 309344, and the companies Knowit, and Iterate through the research projects Transformit and 10XTeams. 
\bibliographystyle{splncs04}
\bibliography{bibliography}

%Author, F.: Article title. Journal \textbf{2}(5), 99--110 (2016)

%Author, F., Author, S.: Title of a proceedings paper. In: Editor,
%F., Editor, S. (eds.) CONFERENCE 2016, LNCS, vol. 9999, pp. 1--13.
%Springer, Heidelberg (2016). \doi{10.10007/1234567890}

%Author, F., Author, S., Author, T.: Book title. 2nd edn. Publisher,

%Author, A.-B.: Contribution title. In: 9th International Proceedings
%on Proceedings, pp. 1--2. Publisher, Location (2010)

%LNCS Homepage, \url{http://www.springer.com/lncs}. Last accessed 4
%\end{thebibliography}

\end{document}